\documentclass[12pt]{article}

\usepackage{amsmath, amsfonts}
\usepackage{graphicx,psfrag,epsf}
\usepackage{enumerate}
\usepackage{natbib}
\usepackage{url} 

\usepackage{titlesec}
\titleformat*{\section}{\normalsize\bfseries}
\titleformat*{\subsection}{\normalsize\bfseries}

\usepackage{comment}
\usepackage{rotating}
\usepackage{amsthm}
\usepackage{amssymb}
\usepackage{setspace}
\usepackage{color}
\usepackage[linesnumbered, ruled]{algorithm2e}
\SetKwRepeat{Do}{do}{while}%

\makeatletter
\newcommand{\vast}{\bBigg@{4}}
\newcommand{\Vast}{\bBigg@{5}}
\makeatother

\def\l{\left}
\def\r{\right}
\def\mc{\mathcal}
\def\bs{\boldsymbol}

\newcommand{\overbar}[1]{\mkern 1.5mu\overline{\mkern-1.5mu#1\mkern-1.5mu}\mkern 1.5mu}
\DeclareMathOperator*{\argmax}{\arg\!\max}
\DeclareMathOperator*{\argmin}{\arg\!\min}
\newcommand{\sign}{\mathrm{sign}}

\usepackage[a4paper]{geometry}

\newtheorem{thm}{Theorem}[section]
\newtheorem{lem}[thm]{Lemma}

\begin{document}

\begin{center}
  \textbf{Estimating Individualized Treatment Regimes from Crossover Designs} \\
	\vspace{0.1in}
  \textbf{Crystal T. Nguyen$^{1}$, 
Daniel J. Luckett$^{1}$, Anna R. Kahkoska$^{2}$,}
  \\
  \textbf{Grace E. Shearrer$^{2}$, Donna Spruijt-Metz$^{3}$, Jaimie N. Davis$^{4}$,} \\
  \textbf{and Michael R. Kosorok$^{1}$} \\
	\vspace{0.1in}
$^{1}$Department of Biostatistics, University of North Carolina, Chapel Hill, North Carolina, U.S.A. \\
$^{2}$Department of Nutrition, University of North Carolina, Chapel Hill, North Carolina, U.S.A. \\
$^{3}$Center of Economic and Social Research, University of Southern California, Los Angeles, California, U.S.A \\
$^{4}$Department of Nutrition, University of Texas at Austin, Austin, Texas, U.S.A.
\end{center}

\normalsize

\thispagestyle{empty}
\begin{abstract}

The field of precision medicine aims to tailor treatment based on patient-specific factors in a reproducible way. To this end, estimating an optimal individualized treatment regime (ITR) that recommends treatment decisions based on patient characteristics to maximize the mean of a pre-specified outcome is of particular interest. Several methods have been proposed for estimating an optimal ITR from clinical trial data in the parallel group setting where each subject is randomized to a single intervention. However, little work has been done in the area of estimating the optimal ITR from crossover study designs. Such designs naturally lend themselves to precision medicine, because they allow for observing the response to multiple treatments for each patient. In this paper, we introduce a method for estimating the optimal ITR using data from a $2\times2$ crossover study with or without carryover effects. The proposed method is similar to policy search methods such as outcome weighted learning (OWL); however, we take advantage of the crossover design by using the difference in responses under each treatment as the observed reward. We establish Fisher and global consistency, present numerical experiments, and analyze data from a feeding trial to demonstrate the improved performance of the proposed method compared to standard methods for a parallel study design.
\end{abstract}

\noindent%

\newpage

\setcounter{page}{1}
\section{Introduction}
Personalized medicine is the practice of tailoring treatment to account for patient heterogeneity \citep{chakraborty2014dynamic}. Physicians and other health care providers have practiced personalized medicine by adjusting doses or prescriptions based on a patient's medical history or demographics for centuries \citep{ashley2015precision, zhao2013recent}. Precision medicine is an emerging field that aims to support personalized medicine decisions with reproducible research \citep{collins2015new}. Such research is imperative, particularly when diseases are expressed with great heterogeneity across patients. A topic of interest in precision medicine is the individualized treatment regime (ITR): a set of decision rules for one or more decision time points that can be used to assign patients to treatment tailored by their patient-specifict factors \citep{lavori2014introduction, moodie2007demystifying, petersen2007individualized}. One objective in precision medicine is to estimate the optimal ITR, or the ITR that maximizes the mean of some desirable outcome \citep{kosorok2015adaptive, laber2014dynamic}. Crossover clinical trials are uniquely suited to precision medicine, because they allow for observing responses to multiple treatments for each patient. This paper introduces a method to estimate optimal ITRs using data from a crossover study by extending generalized outcome weighted learning (GOWL) \citep{chen2018estimating} to deal with correlated outcomes.

In a crossover study, patients are randomized to a sequence of treatments rather than a single treatment. Thus, multiple outcomes are observed, one per subject from each treatment period, and each subject acts as his or her own control for reduced between-subject variability \citep{machin2010randomized, turner2010new, wellek2012proper}. Therefore, crossover designs naturally lend themselves to precision medicine; estimating the optimal ITR from a crossover design can utilize all counterfactual outcomes. In contrast, estimating the optimal ITR from traditional parallel group designs, where patients are assigned to a single treatment, can only utilize the subset of counterfactual outcomes that are observed.

There have been many developments in machine learning methods for answering precision medicine questions from parallel study designs. For example, \cite{qian2011performance} indirectly estimate the decision rule using L1 penalized least squares; Zhang et al. (2012a)\nocite{zhang2012robust} maximize a doubly robust augmented inverse probability weighted estimator for the population mean outcome; \cite{athey2017efficient} maximize a doubly robust score that may take into account instrumental variables; \cite{kallus2018balanced} employs a weighting algorithm similar to inverse probability weighting but minimize the worst case mean square error; \cite{laber2015tree} propose the use of decision trees, which prove to be both flexible and easily interpretable; \cite{zhao2012estimating}, Zhang et al. (2012b)\nocite{zhang2012estimating}, \cite{zhou2017residual}, and \cite{chen2018estimating} directly estimate the decision rule by viewing the problem from a weighted classification standpoint.

However, little work has been done to develop precision medicine methods that handle correlated observations in the single-stage decision setting such as those that arise from crossover designs. \cite{hawaii} propose a weighted ranking algorithm to estimate a decision rule that maximizes either the expected outcome or the probability of selecting the best treatment, but they assume that there are no carryover effects present. Because the intended effect of the washout period can be difficult to achieve in practice \citep{wellek2012proper}, it is imperative that methods for crossover designs can be applied when carryover effects are present. In this paper, we show that the difference in response to two treatments from a $2\times2$ crossover trial can be used as the reward in the generalized outcome weighted learning (GOWL) objective function to estimate an optimal ITR. We introduce a plug-in estimator that can be used with the proposed method to account for carryover effects. Additionally, we show that using a crossover design with the proposed method results in improvements in misclassification rate and estimated value when compared to standard methods for a parallel design at the same sample size.

As a clinical example, consider nutritional recommendations surrounding the intake of dietary fiber for the purpose of weight loss. Although increased fiber is recommended across the population for a myriad of health benefits \citep{anderson1994health, anderson2009health, marlett2002position, us2010us}, evidence of the impact of the consumption of dietary fiber for improved satiety and reduction in body weight is mixed \citep{halliday2018adolescents, slavin2005dietary}. Heterogeneity in response to dietary fiber may be leveraged to develop targeted fiber interventions to promote feelings of satiety. We use data from a crossover study in which Hispanic and African American adolescents who are overweight and obese were fed breakfast and lunch under a typical western high sugar diet and a high fiber diet. From these data, we estimate a decision rule with which clinical care providers can input patient characteristics, including demographics and clinical measures, and receive a recommendation to maximize the change in measures of perceived satiety from before breakfast to after lunch. This type of analysis could be useful in identifying a subgroup of at-risk adolescents for which targeting specific dietary recommendations is expected to lead to an increase in patient-reported satiety, helping to decrease caloric intake in a population with great clinical need for effective weight loss strategies.

The rest of this paper is organized as follows. In Section 2, we review outcome weighted learning (OWL) \citep{zhao2012estimating} and present the proposed method for estimating an optimal ITR from a crossover study regardless of the presence of carryover effects. Section 3 establishes Fisher and global consistency. Section 4 demonstrates the performance of the proposed method in simulation studies, with results on misclassification rate and estimated value. Section 5 reports on the analysis of data from a feeding trial with overweight and obese Latino and African American adolescents, and we conclude with a discussion in Section 6. 

\section{Methodology}
In this section, we provide a brief overview of existing methods for estimating the optimal ITR using weighted classification. We then provide the justification and means to implement our proposed method, which we will from here refer to as ``crossover GOWL."

\subsection{Existing Methods}
Consider a parallel, two-arm clinical trial in which we have i.i.d. observations $(\bs{X}_i, A_i, Y_i)$ for $i=1,\ldots,n$, where $A\in\mc{A}=\{-1,1\}$ is binary treatment assignment, $\bs{X}\in\mc{X}$ is a $p$-dimensional vector of covariates, and $Y\in\mathbb{R}$ is a reward, bounded by $M_0<\infty$, for which greater values are desired. Assume that $Y$ is of the form $$Y = \mu(\bs{X}) + Ac(\bs{X}) + \epsilon,$$ where $\mu(\bs{X})$ is the main effect of the covariates, $c(\bs{X})$ is the treatment-covariate interaction, and $\epsilon$ has mean 0 and variance $\sigma^2_{\epsilon}$. Denote $Y^*(a)$ as the counterfactual outcome under treatment $a$. We then make three causal assumptions \citep{rubin1978bayesian} to connect the counterfactual outcomes to the observed data: $P(A=a|\bs{X})>0$ with probability 1, $\{Y^*(1),Y^*(-1)\}\perp A|\bs{X}$, and $Y=Y^*(a)$. These are known as positivity, conditional exchangeability, and consistency, respectively.

An ITR, $D$, comes from the set of all functions $\mc{D}$ that map the covariate space $\mc{X}$ to the treatment space $\mc{A}$. Our objective is to estimate the optimal ITR, denoted $D_0$, which maximizes the value function \citep{qian2011performance}, \begin{equation}
  \mc{V}(D)=E\l[\frac{Y1\{A=D(\bs{X})\}}{P(A|\bs{X})}\r],
  \label{value1}
\end{equation}
where $P(A|\bs{X}) = \mathrm{Pr}(A=a|\bs{X}=\bs{x})$ is the propensity score for treatment. Equivalently, $D_0$ may be defined as
\begin{equation}
  D_0=\argmin_{D\in\mc{D}}E\l[\frac{Y1\{A\neq D(\bs{X})\}}{P(A|\bs{X})}\r].
  \label{optTx1}
\end{equation}

\cite{zhao2012estimating} propose OWL to solve this problem: each misclassified observation is weighted by its observed outcome, $Y$, and the hinge loss is used to bring the problem into the support vector machine framework \citep{cortes1995machine}. Unfortunately, OWL assumes $Y$ is nonnegative; when negative values are observed, OWL shifts all outcomes to be nonnegative, since (\ref{optTx1}) is invariant to such a transformation. The objective function in OWL, however, does not have this property. Therefore, the estimated decision function in OWL depends on the chosen shift in the outcomes. \cite{chen2018estimating} propose GOWL, an extension of OWL, which handles negative rewards by modifying the hinge loss to be piecewise and weighting the misclassified observations by $|Y|$. With GOWL, there is no need to shift rewards. 

However, neither \cite{zhao2012estimating} nor \cite{chen2018estimating} considered correlated outcomes, such as those that arise from a crossover design setting. We now introduce crossover GOWL, a method that combines the observed treatment response difference with GOWL to estimate the optimal ITR from $2\times2$ crossover data.

\subsection{Crossover Generalized Outcome Weighted Learning}

In a crossover design, patients are randomly assigned to a sequence of treatments rather than a single treatment. For the $2\times2$ design, patients are randomized to receive either the $(-1, 1)$ or the $(1, -1)$ sequence, with some prespecified washout period between treatments. The washout period is a break between treatments which serves to remove any carryover effects, or residual effects remaining from a previous treatment at the start of the next treatment. Keeping most of the notation from before, we now introduce sequential treatments and outcomes $A_k$ and $Y_k$ for periods $k=1,2$, respectively, i.e., $Y_k$ is the observed outcome after receiving treatment $A_k$ in period $k$. Furthermore, we assume the model $$Y_k=\mu(\bs{X})+A_kc(\bs{X})+\delta_{A_1}(\bs{X})\ 1\{k=2\}+\epsilon_k$$ where $\bs{\epsilon}=(\epsilon_1,\epsilon_2)^{\top}$ has mean $\bs{0}$ and a positive definite covariance matrix, $\Sigma_{\epsilon}$, and $\delta_{A_1}(\bs{X})$ is the carryover effect which may depend on $A_1$ and $\bs{X}$. Note that in a $2\times2$ crossover study, the period effects, or temporal effects, are nonseparable from the carryover effects \citep{fleiss1989critique}, so $\delta_{A_1}(\bs{X})$ encompasses both period and carryover effects. 

Let $R = Y_1-[Y_2-\delta_{A_1}(\bs{X})]$. Given the observed data $(\bs{X},A_1,Y)$, we propose the following as a substitute for the value function to be maximized:
\begin{equation}
  E\l[\frac{R}{P(A_1|\bs{X})}1\{A_1=D(\bs{X})\}\r],
  \label{value2}
\end{equation}
where $P(A_1|\bs{X})$ is the probability of being assigned to the sequence $(A_1, -A_1)$ conditional on $\bs{X}$. Under Lemma \ref{lemma}, maximizing (\ref{value2}) is equivalent to maximizing (\ref{value1}); the proof is left to Appendix C.

\begin{lem} Under the given assumptions,
$$D_0=\argmin_{D\in{D}}E\l[\frac{R}{P(A_1|\bs{X})}1\{A_1\neq {D}(\bs{X})\}\r].$$
\label{lemma}
\end{lem}

Following (\ref{value2}), we use an approach similar to GOWL but weight misclassified observations by the treatment response difference, and we minimize the objective function (\ref{eqn:empirical}) for $f$ in $\mc{F}$, a class of functions, e.g., a reproducing kernel Hilbert space. Let $\psi(u,v)=\max\{1-\sign(u)v, 0\}$, $\lambda_n$ be a tuning parameter, and $||f||$ be the $L_2$ norm of $f$. For details on solving the minimization problem in (\ref{eqn:empirical}), we defer to \cite{chen2018estimating} and \cite{kimeldorf1970correspondence}. 
\begin{equation}
  \argmin_{f\in\mc{F}}\frac{1}{n}\sum_{i=1}^n\frac{|R_i|}{P(A_{i, 1}|\bs{X}_i)}\psi\{R_i,A_{i,1}f(\bs{X}_i)\}+\lambda_n||f||^2,
  \label{eqn:empirical}
\end{equation}

In practice, the true value of $\delta_{A_1}(\bs{X})$ is unknown. In traditional analyses, we are concerned with testing the null hypothesis that $\delta_{-1}(\bs{X})=\delta_1(\bs{X})$. Here, we are instead interested in whether or not either treatment has a nonzero carryover effect. Investigators may determine whether carryover effects are present any number of ways, including two-sample $t$-tests for the null hypotheses $H_{0,1}{:}\ E[\delta_1(\bs{X})]=0$ and $H_{0,-1}{:}\ E[\delta_{-1}(\bs{X})]=0$ by comparing mean responses to each treatment at each time point. An estimator for $\delta_{A_1}(\bs{X)}$, denoted $\widehat{\delta}_{A_1}(\bs{X})$, can be computed using Algorithm 1.

\IncMargin{1em}
\begin{algorithm}
 \caption{Estimating $\delta_{A_1}(\bs{X})$}
 Estimate $g(\bs{x},a_1)=E[Y_1|\bs{X}=\bs{x},A_1=a_1]$ by regressing $Y_1$ on $\bs{X}$ and $A_1$.\\
  Set $\widehat{Y}_2=\widehat{g}(\bs{X},A_2)$. \\
  Estimate $\delta_{A_1}(\bs{X})$ by regressing $Y_2-\widehat{Y}_2$ on $\bs{X}$ and $A_1$. 
\end{algorithm}
\DecMargin{1em}

In short, one model is fit to predict what would have been observed in period 2 in the absence of carryover effects, and another model is fit to predict the residual from the first model. While any regression technique may be used here, we use reinforcement learning trees (RLT) in our implementation. RLT is a nonparametric tree-based machine learning method that considers future splits or branches in the model when determining the best split at any node \citep{zhu2015reinforcement}.

We can now correct the observed reward with the estimated carryover effects. Letting $\widehat{R} = Y_1 - \left[Y_2 - \widehat{\delta}_{A_{i,1}}(\bs{X})\right],$ the estimated decision function is
\begin{equation}
  \widehat{f}^*_n=\argmin_{f\in\mc{F}}\frac{1}{n}\sum_{i=1}^n\frac{|\widehat{R}_i|}{P(A_{i, 1}|\bs{X}_i)}\psi\left\{\widehat{R}_i,A_{i,1}f(\bs{X}_i)\right\}+\lambda_n||f||^2,
  \label{eqn:carryover}
\end{equation}
and our proposed estimator of the optimal ITR is $\widehat{D}^*(\bs{X}) = \sign\left\{\widehat{f}^*_n(\bs{X})\right\}$, where $$D^*=\argmax_{D\in\mc{D}}\ E\left[\frac{|R|}{P(A_1|\bs{X})}\psi\{R,A_1f(\bs{X})\}\right].$$ 

\section{Theoretical Results}

In this section, we establish both Fisher and global consistency. First, define the risk under 0-1 loss to be $\mc{R}(f)=E[Y\ 1\{A\neq\sign[f(\bs{X})]\}/P(A|\bs{X})].$ The risk under the modified loss function with the reward defined as the treatment response difference is then $$\mc{R}_{\psi}(f)=E\l[\frac{|\widehat{R}|}{P(A_1|\bs{X})}\psi\left\{\widehat{R},A_1f(\bs{X})\right\}\r].$$ Let $f^*(\bs{X})=\argmin_{f\in\mc{F}}\mc{R}_{\psi}(f)$, so that the corresponding ITR under the modified loss for the treatment response difference is ${D}^*(\bs{X})=\sign\{f^*(\bs{X})\}$. Under Theorem 1, Fisher consistency for $D^*(\bs{X})$ is derived.

\begin{thm}
Under the given assumptions, ${D}^*(\bs{X})={D}_0(\bs{X})$.
\label{t1}
\end{thm}

Consider that $\mc{F} = \{k(\cdot, \bs{x}): \bs{x}\in\mc{X}\}$ for some kernel function $k$, and let $\overbar{\mc{F}}$ be the closure of $\mc{F}$. Define $f_0$ to be the minimizer over all functions $f$ for $\mc{R}(f)$, and define $f_0^*$ to be the same for $\mc{R}_{\psi}(f)$. 

\begin{thm}
Let $\lambda_n>0$ be a sequence such that $\lambda_n\to0$ and $\lambda_nn\to\infty$ with probability going to 1 as $n\to\infty$. Assume $\exists\ M_1<\infty$ such that $P\left(|\widehat{\delta}_{A_1}(\bs{X})|<M_1\right)\to1$ as $n\to\infty$ and $|\delta_{A_1}(\bs{X})|<M_1$ almost surely. If $\mathbb{P}\left[1\{\sign[\widehat{R}]\neq\sign[R]\}\right]=o_P(\lambda_n)$, then, for any distribution $P$ of $(\bs{X}, A_1, \bs{Y}),$ $\lim_{n\to\infty}\mc{R}_{\psi}(\widehat{f}^*_n)\to_P\mc{R}_{\psi}\left(f^*\right)$. Furthermore, if $f^*_0\in\overbar{\mc{F}},$ $$\lim_{n\to\infty}\mc{R}(\widehat{f}^*_n)\to_P\mc{R}\left(f_0\right).$$ 
\label{t2}
\end{thm}
Derivation of Theorems 1 and 2 may be found in Appendix C.

\section{Simulation Studies}
To illustrate the benefits of using crossover GOWL, we present simulation studies with comparisons to standard methods used in parallel group clinical trials. Simulated data sets were generated as follows. The covariates, $\bs{X}_1,\ldots,\bs{X}_{50}$, are i.i.d. variables drawn from a $U(-1, 1)$ distribution. Subjects were randomized to treatment $-1$ or $1$ for the parallel design or to sequence $(-1, 1)$ or $(1, -1)$ for the crossover design with equal probability. The response for the parallel design, $Y,$ is normally distributed with a mean of $\mu(\bs{X})+c(\bs{X})A$ and a variance of 1. For the crossover design, responses were simulated per the model $Y_k=\mu(\bs{X})+A_kc(\bs{X})+\delta_{A_1}(\bs{X})\ 1\{k=2\}+\epsilon_k$, for $k=1,2$, where $\bs{\epsilon}$ was drawn from a multivariate normal distribution with mean $\bs{0},\ \mathrm{Var}[\epsilon_1]=\mathrm{Var}[\epsilon_2]=1$, and $\mathrm{Cov}[\epsilon_1,\epsilon_2]=0.5$. $\mu(\bs{X})$ was fixed to be $1 + \bs{X}_1 + 2\bs{X}_2 + 0.5\bs{X}_3+\bs{X}_4$ for all simulation scenarios. Table \ref{tab:simscenarios} describes choices of $c(\bs{X})$ and $\delta_{A_1}(\bs{X})$ defining four scenarios. 

\begin{table}[h!]
  \centering
  \caption{The interactive and carryover effects for the five simulation scenarios.}
  \begin{tabular}{clcc}
     \hline 
     Scenario & $c(\bs{X})$ & $\delta_{-1}(\bs{X})$ & $\delta_1(\bs{X})$ \\
     \hline 
     1 & $1.12(0.3-X_1-X_2)$ & 0 & 0\\ 
     2 & $1.15(X_1-1.25X_2^2)$ & 0 & 0\\
     3 & $1.12(0.3-X_1-X_2)$ & $\l|\frac{\mu(\bs{X})+c(\bs{X})}{4}\r|$ & $\l|\frac{\mu(\bs{X})-c(\bs{X})}{2}\r|$\\
     4 & $1.15(X_1-1.25X_2^2)$ & $0.4X_1^2+0.3X_2$ & $1-2X_1-X_2^2$\\
     \hline
  \end{tabular}
  \label{tab:simscenarios}
\end{table}

Scenarios 1 and 3 are linear in $\bs{X}$, whereas Scenarios 2 and 4 are nonlinear. Note that scenario pairs $(1, 3)$ and $(2, 4)$ are similar, but Scenarios 3 and 4 include carryover effects. The optimal ITR was estimated via crossover GOWL, using a Gaussian kernel. The penalty parameter, $\lambda_n,$ and the Gaussian kernel bandwidth parameter, $\sigma_n,$ were selected using 5-fold cross-validation on the grids $\{0.1, 0.5, 1, 5, 10, 50, 100, 500\} / n$ and $(0.1, 0.2,\ldots, 5.0)$, respectively. In scenarios where carryover effects are present, RLT \citep{zhu2015reinforcement} was used to fit both models to estimate $\widehat{\delta}_{A_1}(\bs{X})$ using Algorithm 1.

A testing data set of size $n_{\mathrm{test}}=10,000$ was generated similarly with period 1 data only. The misclassification rate, or $\mathbb{P}_{n_{\mathrm{test}}}1\left\{\widehat{{D}}^*(\bs{X})\neq{D}_0(\bs{X})\right\}$, of the estimated ITR applied to the testing set was calculated, where $\mathbb{P}_{n_{\mathrm{test}}}$ is the empirical mean in the test set. We also calculated the estimated value of the estimated ITR, $\widehat{\mc{V}}\left(\widehat{D}^*\right)$ \citep{qian2011performance}, where
\begin{equation}
\widehat{\mc{V}}(D)=\frac{\mathbb{P}_{n_{\mathrm{test}}}[Y1\{A=D(\bs{X})\}/P(A_1|\bs{X})]}{\mathbb{P}_{n_{\mathrm{test}}}[1\{A=D(\bs{X}\}/P(A_1|\bs{X})]}.
\label{eqn:estvalue}
\end{equation} 
Note that $P(A_1|\bs{X})=0.5$ is constant here. The estimated value is the average reward observed under the estimated optimal ITR when applied to the testing set. Figure 3 in Appendix A displays the mean square error from estimating the carryover effects with RLT for Scenarios 3 and 4.

Simulations were repeated 1,000 times at training set sample sizes of 30, 75, 150, 300, and 600. Comparisons to various methods in the parallel setting at the same sample size are presented in Figures 1 and 2. These methods include OWL, GOWL, and ridge regression. For OWL and GOWL, a Gaussian kernel was used, and the aforementioned grids for $\lambda_n$ and $\sigma_n$ are considered in 5-fold cross-validation. 
For ridge regression, the model includes all covariates and treatment-covariate interactions without any higher order terms or between-covariate interactions. 5-fold cross-validation was used to determine a value for the the ridge penalty parameter, where the same values for $\lambda_n$ in the OWL methods are considered. All simulations were performed with R version 3.4.3 \citep{R}. RLT was implemented with the RLT package, version 3.2.1 \citep{RRLT}, and all OWL methods were implemented with the DynTxRegime package, version 3.2 \citep{RDyntxregime}. While the DynTxRegime package does not currently support GOWL, the inputs for OWL can be recoded to implement GOWL. Ridge regression was carried out with the glmnet package \citep{glmnet}. 

\begin{figure}[h!]
  \centering
  \includegraphics[width=\textwidth]{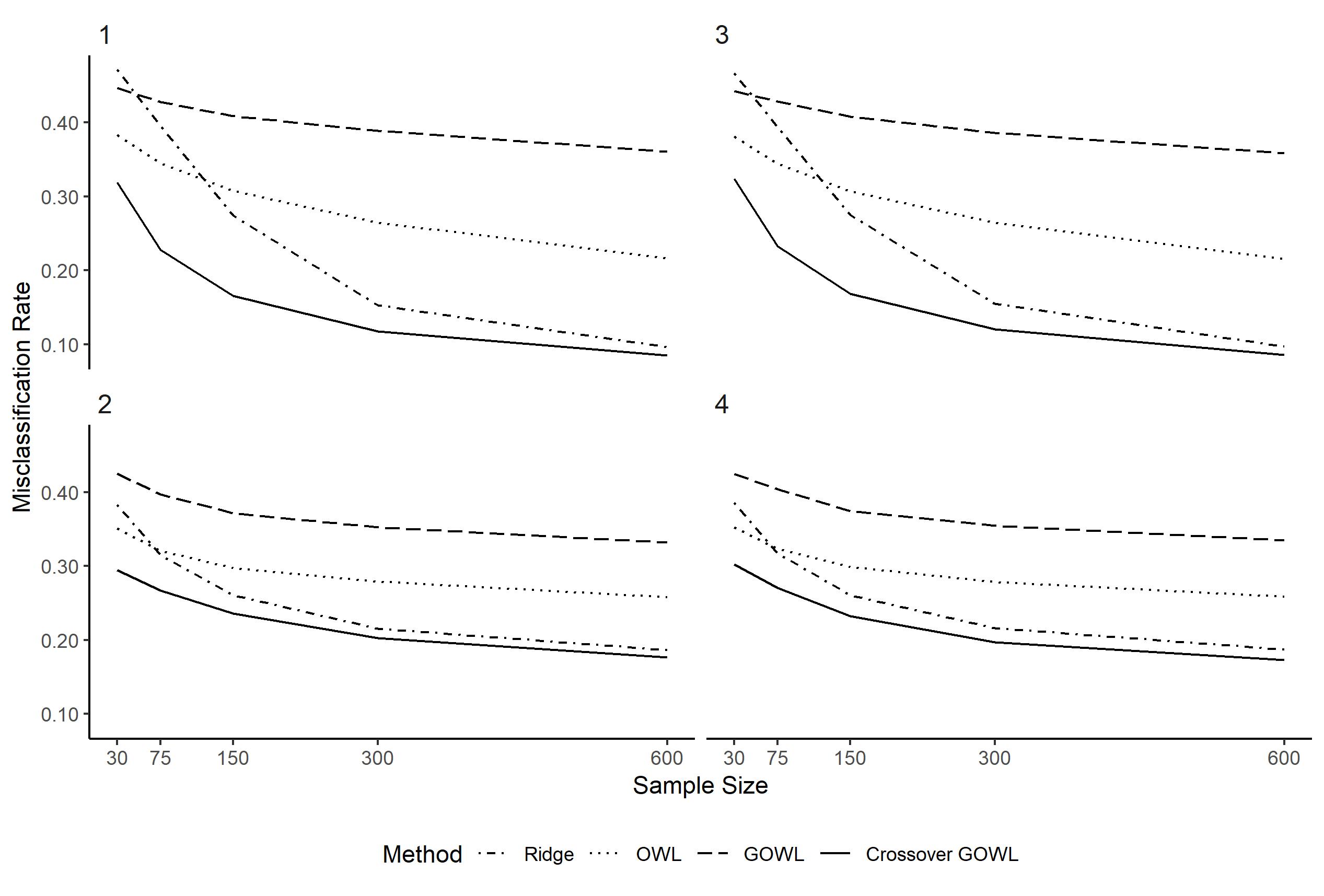}
  \caption{Mean missclassification rate of 1,000 simulations for estimating the optimal ITR, applied to a testing set of size 10,000 for each of 4 simulation scenarios.}
  \label{fig:misclass}
\end{figure}
\begin{figure}[h!]
  \centering
  \includegraphics[width=\textwidth]{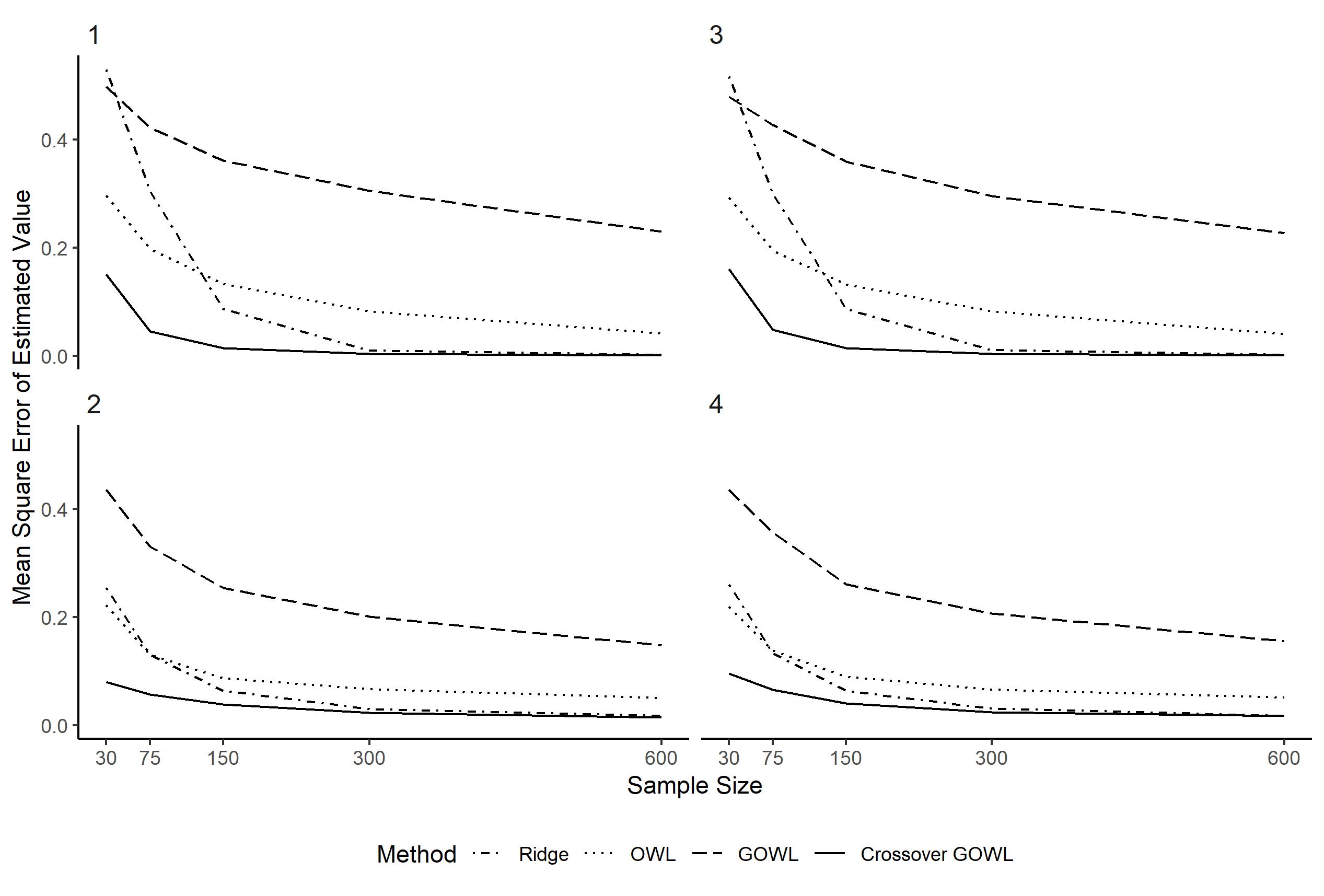}
  \caption{Mean square error of the estimated value compared to the true value from 1,000 simulations for estimating the optimal ITR, applied to a testing set of size 10,000 for each of 4 simulation scenarios.}
  \label{fig:value}
\end{figure}

Figure \ref{fig:misclass} displays the average misclassification rates across all sample sizes, methods, and scenarios. Figure \ref{fig:value} displays the mean square error of the estimated value from the true value, i.e., $\mathbb{P}_{n_{\mathrm{test}}}\left\{\left[\widehat{\mathcal{V}}\left(\widehat{D}^*\right)-\widehat{\mathcal{V}}\left(D_0\right)\right]^2\right\}$ from period 1 data. On average, crossover GOWL yields lower misclassification rates and higher estimated values at smaller sample sizes across all scenarios. Crossover GOWL shows marked improvement in both misclassification and estimated value for small $n$. When $n$ is large, ridge regression yields competitive results with that from crossover GOWL, but crossover GOWL still appears to have marginal gains. Although GOWL in the parallel setting does not perform as well as OWL in any of the presented scenarios, \cite{chen2018estimating} discuss scenarios where improvements in misclassification and estimated value are observed when using GOWL as opposed to OWL. 

\section{FAME Feeding Trial Data Analysis}

We present the application of crossover GOWL to data from the Food, Adolescents, Mood and Exercise (FAME) crossover feeding trial, conducted at the University of Southern California (USC) \citep{o2015effects}. The FAME trial included African American and Latino adolescents who were overweight or obese. African American and Latino adolescents are disproportionately affected by overweight and obesity outcomes compared to their non-Hispanic counterparts \citep{ogden2014prevalence, o2015effects, taveras2013reducing}. Dietary intake is a major modifiable risk factor and represents a key intervention point in improving weight loss \citep{bleich2017interventions, kipping2008obesity}. One promising approach is to modify dietary components to improve satiety to indirectly reduce caloric intake \citep{anderson2009health}. In epidemiologic studies of adults in the US, fiber intake is inversely associated with body weight and body fat \citep{slavin2005dietary}, even after adjusting for confounding factors such as dietary fat intake. However, results from intervention studies are mixed: increased dietary fiber intake has been shown to have varied effects on body weight among adults who are overweight or obese, with limited research in pediatric or adolescent populations \citep{rossner1987weight, ryttig1989dietary, slavin2005dietary, thompson2005effect, tucker2009increasing}. Given the heterogeneity in the effects of dietary fiber intake on body weight, it is essential to identify the subgroups of overweight and obese adolescents who may benefit from tailored clinical advice to increase fiber intake. We estimate a decision rule to identify a subgroup of adolescents who are overweight or obese that experiences larger increases in patient-reported satiety from a high fiber diet as opposed to the more common high sugar diet.

This study was conducted at the USC Health Sciences campus in Los Angeles, California from 2008 to 2011. Eighty-six Latino and African American adolescents (ages 14 to 17 years of age) who were overweight or obese (BMI percentile $>85\%$) were recruited. Race was self-reported, and subjects were included if all four grandparents were Latino or African American. Subjects were excluded if they had type 2 diabetes, were in a weight loss program within the past 6 months, or used medications that influenced insulin or body composition. Informed written parental consent and participant assent were acquired before all testing procedures. The Institutional Review Board of USC approved all study procedures. 

Participants received either a high sugar/low fiber (HSLF) meal plan or a high fiber/low sugar (HFLS) meal plan for breakfast and lunch on two separate visit days. Participants were randomized with equal probability to receive the HSLF/HFLS or HFLS/HSLF sequence with a minimum 2 week washout period between visits. The meals were isocaloric and matched for macronutrients except sugar and fiber content. Participants initially attended a baseline visit at the Clinical Trials Unit at the USC University Hospital where insulin sensitivity, Tanner stage via examination by a medical professional, BMI percentile for age, sex, ethnicity, waist circumference, and hemoglobin A1c (HbA1c) were collected. Insulin  sensitivity was assessed via a frequently sampled intravenous glucose tolerance test (FSIVGTT) and calculated using the minimal model \citep{bergman1979quantitative, yang1987modified}. At the subsequent test meal visits, participants received either a HSLF or HFLS breakfast after a 10 hour overnight fast. At noon, the participants received the same meal condition for lunch. Participants rated their hunger and fullness via a 100 mm-visual analog scale (VAS) prior to breakfast and 45 minutes after the start of lunch (300 minutes after breakfast). Participants were provided with age appropriate activities between meals (e.g., video games, crafts, books, etc.).  

The satiety outcomes are formally defined as the negative change in hunger, since lower values of hunger are desired, and the observed change in fullness between 8:00 AM and 1:00 PM (before breakfast and after lunch). Due to the nature of the outcomes, the required 10 hour overnight fast, and the implemented minimum 2 week washout period, we assumed no carryover effects were present. Of the 86 subjects who completed the study, 20 were removed for missing outcomes, and 1 was removed for missing insulin sensitivity. Participants that did not return within 5 weeks were also removed $(n=54)$. We compared crossover GOWL with OWL, GOWL, and ridge regression using data from period 1 only. Methods were implemented as described in Section 4. 5-fold cross-validated value estimates were obtained, but rather than using Equation (\ref{eqn:estvalue}) which uses only period 1 data, the value for each observation $i=1,\ldots,n_m$ in the $m$th fold's testing set was computed as $Y_{i,1}1\l\{A_1=\widehat{D}_0(\bs{X})\r\}+Y_{i,2}1\l\{A_2=\widehat{D}_0(\bs{X})\r\}$ where $n_m$ is the size of the $m$th fold for $m=1,\ldots,5.$ Although OWL, GOWL, and ridge regression were trained on period 1 data, data from both periods were used to improve accuracy in the value estimate because the testing set size for each fold is quite small.

Resulting estimated values, averaged across folds, are presented in Table \ref{tab:realdata} along with the mean outcome observed from period 1. For both outcomes, all methods show improvement in the estimated value in comparison to randomization, but crossover GOWL yields the highest improvement. For self-reported fullness, crossover GOWL also yields the smallest standard deviation. When training crossover GOWL on the full dataset, 92\% (51\%) of participants are assigned to the HFLS to maximize the change in fullness (hunger). The distribution of features across the groups assigned to HFLS and HSLF from crossover GOWL for both outcomes are presented in Figure 4 in Appendix B. Dietary fiber is recommended to improve overall health in the general population \citep{marlett2002position}; however, the estimated ITRs from hunger and fullness may inform the development of tailored dietary intake advice for subgroups of at-risk adolescents.

\begin{table}[h!]
  \centering
  \caption{Mean (sd) 5-fold cross-validated estimated values for feeding trial data compared with the observed value from period 1.}
  \begin{tabular}{lrrrr}
  \hline & \multicolumn{4}{c}{Outcome} \\
  \cline{2-5}
   & \multicolumn{2}{c}{Fullness} & \multicolumn{2}{c}{Hunger} \\
  \hline 
  Ridge & $3.00$ & $(4.53)$ & $5.60$ & $(8.15)$ \\
  OWL & $3.07$ & $(3.88)$ & $5.45$ & $(7.42)$\\
  GOWL & $3.85$ & $(4.97)$ & $8.29$ & $(7.93)$\\
  Crossover GOWL & $6.39$ & $(3.57)$ & $10.50$ & $(8.36)$\\
  \hline
  Observed & \multicolumn{2}{c}{$0.96$} & \multicolumn{2}{c}{$4.66$}\\
  \hline
  \end{tabular}
  \label{tab:realdata}
\end{table}

\section{Discussion}

Precision medicine is an emerging field with rapid developments in analytical methods; however, these advancements typically revolve around parallel designs. This paper proposes the combined use of crossover designs and generalized outcome weighted learning for the purpose of estimating optimal ITRs. The proposed method addresses a key gap in the literature; little to no work has been done to better involve crossover designs in precision medicine, despite how naturally crossover studies lend themselves to the field. \cite{hawaii} propose a ranking method to estimate the optimal ITR from a crossover study but provide no recommendations on how to deal with carryover effects. In contrast, crossover GOWL is able to handle such effects. Furthermore, regardless of the presence of carryover effects, the proposed method shows improvements in the estimated value and misclassification rate, especially at the smaller sample sizes typical of crossover designs compared to standard methods with the parallel group design.

An alternative to GOWL that has been developed is residual weighted learning (RWL) \citep{zhou2017residual}. RWL is an extension of OWL that weights the misclassification error by residuals from a model fit to the outcome instead of the observed rewards themselves. Unlike GOWL, RWL uses a non-convex loss function that does not guarantee global minimization \citep{tao2005dc}. In the proposed method, there is no need to include residuals in the weight, because the residuals would cancel when taking the difference between responses to each treatment. Thus, the proposed method avoids specifying a model for the main effect of the covariates.

We note that when the distribution of $\widetilde{A_1}=\sign\{R\}A_1$ is poorly allocated, the cross-validation mechanism for estimating $\lambda_n$ and $\sigma_n^2$ may fail. If there is prior knowledge on the distribution of $\sign\{R\}$, investigators could adjust randomization probabilities when assigning patients to treatment sequences accordingly. Otherwise, it is possible for a training set to not observe at least one $\widetilde{A}_1=1$ or $\widetilde{A}_1=-1$. Lastly, there may be low power in testing $H_0:E[\delta_{A_1}(\bs{X})]=0$ at smaller sample sizes.

Several extensions of estimating the optimal ITR from crossover data are yet to be explored. For example, only the $2\times2$ design was studied in this paper. For larger design schemes, the proposed method could be implemented in a series of binary classifiers as in \cite{dietterich1994solving}. Alternatively, one could expand crossover GOWL to multi-category classification. There have been several developments in multi-category SVM \citep{lee2004multicategory, zhu20041}. More recently, \cite{OWDL} propose an outcome weighted deep learning method to estimate the optimal ITR for multiple treatments. Another possible extension is to consider the residual from modeling the treatment response difference as the observed reward. \cite{fu2016estimating} and \cite{zhou2017residual} have seen favorable results using residual weights, but further improvements may come from using the residuals in outcome weighted learning with the piece-wise hinge loss from GOWL. Finally, the proposed method could be improved upon with methods for variable selection. For example, the $L_1$ penalty could be imposed during optimization to simultaneously restrict model complexity and perform variable selection as suggested by \cite{chen2018estimating}, \cite{song2015sparse}, \cite{xu2015regularized}, and \cite{zhou2017residual}.

\section{Acknowledgments}

The authors were supported in part by NCI P01 CA142538 and NCMHD P60 MD002254-01.

\bibliographystyle{Chicago}
\bibliography{crossover}

\section*{Appendix}

\subsection*{Appendix A}

\begin{figure}[h!]
  \centering
  \includegraphics[width=\textwidth]{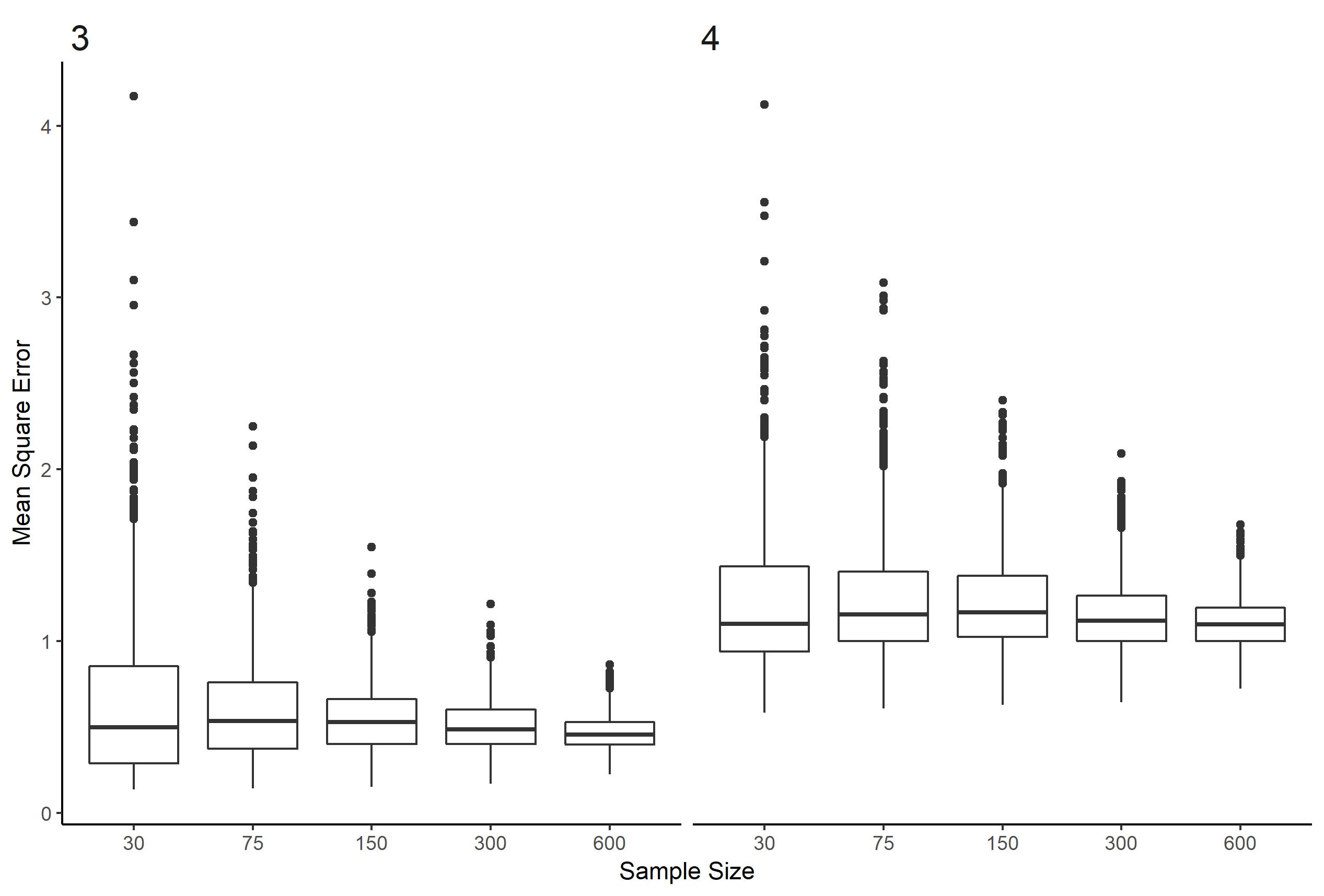}
  \caption{Boxplots of mean square error for RLT predicted $\delta_{A_1}(X)$ on the testing set compared with the true carryover.} 
  \label{fig:mse}
\end{figure}

The reinforcement learning trees (RLT) \citep{zhu2015reinforcement} performance from simulation Scenarios 3 and 4 as visualized in Figure \ref{fig:mse}, which displays the mean square prediction error of the estimated carryover compared with the true carryover from the testing set, or $$\mathbb{P}_{n_{\mathrm{test}}}\left[\widehat{\delta}_{A_{i, 1}}(\bs{X}_i) - \delta_{A_{i, 1}}(\bs{X}_i)\right]^2.$$ Despite the potential for a high mean square error, the crossover design still outperforms parallel design counterparts despite the presence of carryover effects, as can be seen in Figures 1 and 2. 

\subsection*{Appendix B}

92\% of study participants, $(n=49)$ were assigned to the HFLS diet according to crossover GOWL to maximize change in fullness from baseline. To characterize the subgroup that, on average, experiences a larger increase in patient reported fullness, Figure 4 displays the distribution of continuous features across the estimated subgroups. Those assigned to the HFLS diet tend to be older with higher A1c. Because the HSLF group is small ($n=4$), two-sample $t$-tests would not be appropriate to test for significant differences between groups, and trends observed in Figure 4 should be confirmed in future studies.  However, sex $(p=0.1131)$, ethnicity $(p=1)$, and Tanner stage $(p=0.4427)$ were tested using Fisher's exact tests. All tests were non significant at the 0.05 level. 

\begin{figure}[h!]
    \centering
    \includegraphics[width=\textwidth]{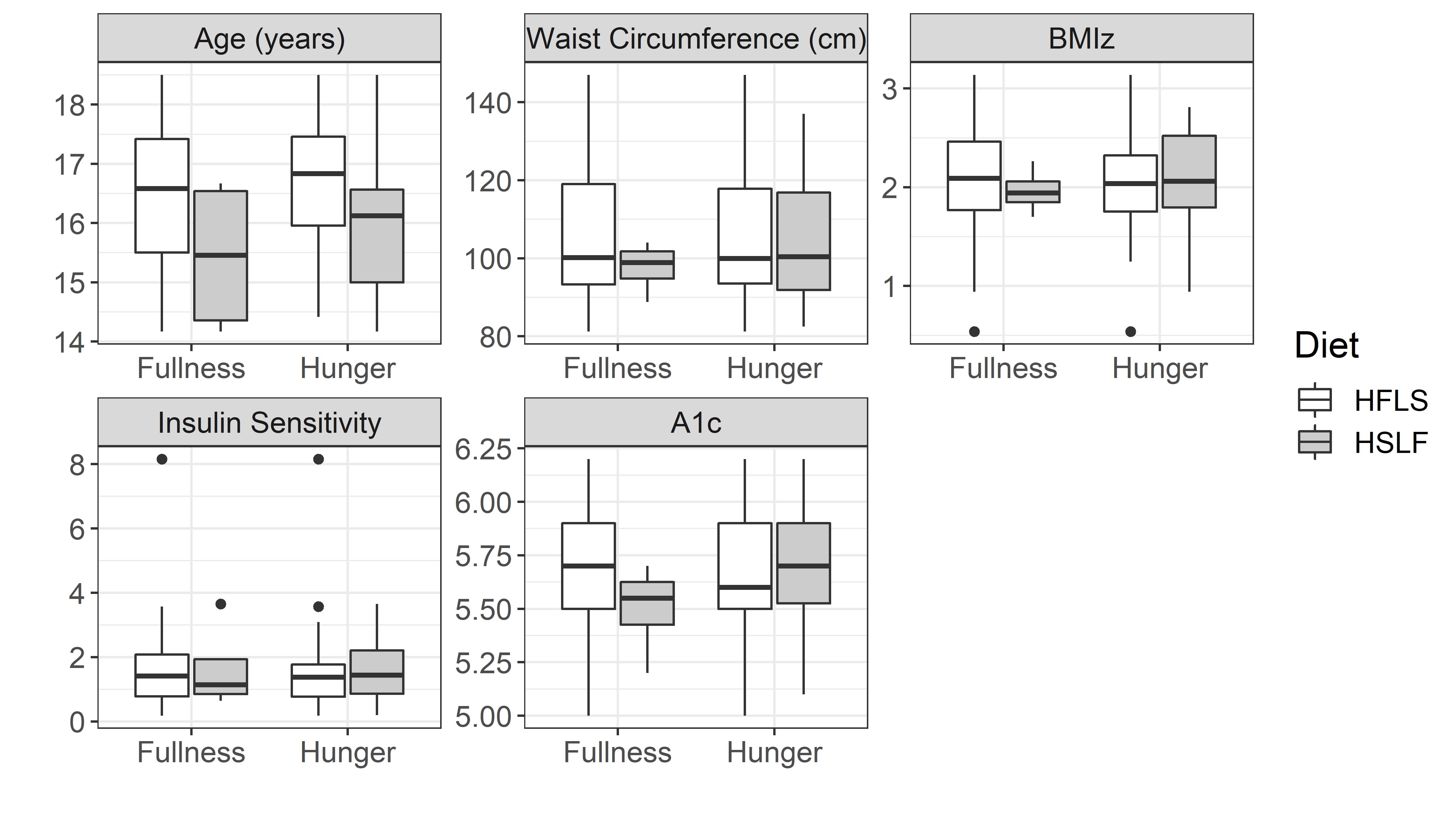}
    \caption{Distribution of features across the crossover GOWL estimated diet-outcome subgroups for both fullness and hunger outcomes}
    \label{fig:features}
\end{figure}

Figure 4 also displays the distribution of continuous features across the estimated subgroups to minimize the change in hunger from baseline. 51\% $(n=27)$ of participants were assigned to HFLS. Those assigned to HFLS tend to be older, but differences in other covariates are not apparent. Although the sample in the HSLF is larger when we consider hunger as the outcome, both samples are still rather small. For this reason, two-sample $t$-tests are still not appropriate. Fisher's exact tests again did not yield any significant differences in sex $(p=0.5857)$, ethnicity $(p=1)$, or Tanner stage $(p=0.7040)$. 

In conclusion, using crossover GOWL appears to be effective for estimating the optimal ITR to maximize the change in satiety. Future research should confirm these subgroups in large sample sizes to better compare differences across features. If verified, future recommendations for adolescent minorities can be tailored by age and A1c levels to improve weight loss. Studies on overweight and obese minority adolescents are still needed to research alternative interventions for those that report feeling more satiated from the typical Western diet (HSLF). 

\subsection*{Appendix C}

The following assumptions are made for the theory behind the method proposed in the main paper.
\begin{enumerate}
  \item \textit{Positivity}: $P(A_1=a|\bs{X}=\bs{x}) \geq\pi_0> 0$ with probability 1 
  \item \textit{Conditional Exchangeability}: $\{Y^*(-1), Y^*(1)\}\perp A_1|\bs{X}$
  \item \textit{Consistency}: $Y_k=Y^*(A_k)-\delta_{A_1}(\bs{X})1\{k=2\}$
  \item Outcomes follow the model $$Y_k=\mu(\bs{X})+A_kc(\bs{X})+\delta_{A_1}(\bs{X})1\{k=2\}+\epsilon_k,$$ for periods $k=1, 2$. $\bs{\epsilon}=(\epsilon_1,\epsilon_2)^{\top}$ has a positive definite covariance matrix, $\Sigma_{\epsilon}.$
  \item There exist $M_0, M_1<\infty$ such that $|Y_k|<M_0$ almost surely, $|\delta_{A_1}(\bs{X})|<M_1$ almost surely, and $P\left(|\widehat{\delta}_{A_1}(\bs{X})|<M_1\right)\to1$ as $n\to\infty$
  \item $\mathbb{P}\left[1\{\sign[\widehat{R}]\neq\sign[R]\}\right]=o_P(\lambda_n)$
\end{enumerate}

\vspace{\baselineskip}

\noindent\textit{Proof of Lemma 1.} The optimal ITR is $D_0=\argmax_{D\in\mc{D}}E[Y^*\{D(\bs{X})\}]$. Note that, under Assumption (4), $D_0(\bs{X})=\sign\{c(\bs{X})\}.$ The expected treatment response difference between treating according to $D_0$ and treating opposite to $D_0$ is

\begin{equation}
\begin{aligned}
E\left[Y^*\{D_0(\bs{X})\}-Y^*\{-D_0(\bs{X})\}\right] &= E[Y^*\{\sign[c(\bs{X})]\}-Y^*\{-\sign[c(\bs{X})]\}] \\
&= 2|c(\bs{X})|\\
&\geq E[Y^*\{D(\bs{X})\}-Y^*\{-D(\bs{X})\}],\nonumber
\end{aligned}
\end{equation}
for all $D\in\mc{D}$. Thus, the optimal ITR also maximizes the treatment-response difference, or $D_0=\argmax_{D\in\mc{D}}E[Y^*\{{D}(\bs{X})\}-Y^*\{-{D}(\bs{X})\}]$. Therefore, it can be seen that
\begin{equation}
\begin{aligned}
D_0&=\argmax_{D\in\mc{D}} E\l[\{Y^*(1)-Y^*(-1)\}D(\bs{X})\r]\\
&= \argmax_{D\in\mc{D}} E\left[\frac{1\{A_1=D(\bs{X})\}}{P(A_1|\bs{X})}\ [Y_1-Y_2+\delta_{A_1}(\bs{X})]\right.\\
&\hspace{0.75cm}+ \left.\frac{1\{A_1\neq D(\bs{X})\}}{P(A_1|\bs{X})} [Y_2-\delta_{A_1}(\bs{X})-Y_1]\right] \\
&= \argmin_{D\in\mc{D}} E\l[\frac{Y_1-Y_2+\delta_{A_1}(\bs{X})}{P(A_1|\bs{X})}\ 1\{A_1\neq D(\bs{X})\}\r] \\ 
&= \argmin_{D\in\mc{D}}E\l[\frac{R}{P(A_1|\bs{X})}\ 1\{A_1\neq D(\bs{X})\}\r]\nonumber,
\end{aligned}
\end{equation}

where the second equality follows from Assumption (3). This proves the result.

\vspace{\baselineskip}

\noindent\textit{Proof of Theorem 1.} This proof follows from Lemma 1 and the results from \cite{lin2002support}. Recall that $\psi(u,v)=\max\{1-\sign(u)v, 0\}$. Minimizing the risk, $\mc{R}_{\psi}(f)$ is equivalent to minimizing the conditional risk, $$\mc{R}_{\psi}(f,\bs{x})=E\left[\frac{|{R}|}{P(A_1|\bs{X})}\psi\{{R},A_1f(\bs{X})\}\Big|\bs{X}=\bs{x}\right],$$ for every fixed $\bs{x}\in\mc{X}$. Let $R^+=R1\{R\geq0\}$ and $R^-=R1\{R<0\}.$ By the law of total expectation, the conditional risk becomes
\begin{equation}
\begin{aligned}
\mc{R}_{\psi}(f,\bs{x})&=E\l[|{R}|\ \psi\{{R},f(\bs{X})\}\Big|\bs{X}=\bs{x},A_1=1\r]+ E\l[|{R}|\ \psi\{{R},-f(\bs{X})\}\Big|\bs{X}=\bs{x},A_1=-1\r]\\
&=E\l[R^+\max\{1-f(\bs{X}),0\}-R^- \max\{1+f(\bs{X}),0\}\Big|\bs{X}=\bs{x},A_1=1\r]\\
&\hspace{0.75cm}+E\l[R^+\max\{1+f(\bs{X}),0\}-R^-\max\{1-f(\bs{X}),0\}\Big|\bs{X}=\bs{x},A_1=-1\r].\nonumber
\end{aligned}
\end{equation}

Next, note that $\mc{R}_{\psi}\{\sign(f),\bs{x}\}<\mc{R}_{\psi}(f,\bs{x})$ whenever $f(\bs{x})\not\in[-1,1].$ For example, when $f(\bs{x})<-1$, the conditional risk reduces to $$[1-f(\bs{x})]\left\{E\left[R^+\Big|\bs{X}=\bs{x},A_1=1\right]-E\left[R^-\Big|\bs{X}=\bs{x},A_1=-1\right]\right\},$$ which is monotonically increasing as $f(\bs{x})\to-\infty$. A similar argument is made for when $f(\bs{x})>1$. Thus, we restrict our search to $f(\bs{x})\in[-1,1]$. Then,

\begin{equation}
\begin{aligned}
\mc{R}_{\psi}(f,\bs{x})&=E\left[R^+-R^-\Big|\bs{X}=\bs{x},A_1=1\right]+E\left[R^+-R^-\Big|\bs{X}=\bs{x},A_1=-1\right]\\
&\hspace{0.75cm}f(\bs{X})\left\{-E\left[R^+-R^-\Big|\bs{X}=\bs{x},A_1=1\right]+E\left[R^++R^-\Big|\bs{X}=\bs{x},A_1=-1\right]\right\}\\
&=E\left[|R|\Big|\bs{X}=\bs{x},A_1=1\right]+E\left[|R|\Big|\bs{X}=\bs{x},A_1=-1\right]\\
&\hspace{0.75cm}+f(\bs{X})\left\{E\left[R\Big|\bs{X}=\bs{x},A_1=-1\right]-E\left[R\Big|\bs{X}=\bs{x},A_1=1\right]\right\}.\nonumber
\end{aligned}
\end{equation}

If $f^*(\bs{x})$ minimizes the conditional risk, then $f^*(x)$ must have the sign opposite of the expression $E\left[R\Big|\bs{X}=\bs{x},A_1=-1\right]-E\left[R\Big|\bs{X}=\bs{x},A_1=1\right]$, and thus $D_0(\bs{X})=\sign\{f^*(\bs{X})\}$

\vspace{\baselineskip}

\noindent\textit{Proof of Theorem 2.} First, define the loss functions $$L_{\psi}(f)=\frac{|R|}{P(A_1|\bs{X})}\psi\{R,A_1f(\bs{X})\}$$ and $$\widehat{L}_{\psi}(f)=\frac{\left|\widehat{R}\right|}{P(A_1|\bs{X})}\psi\{\widehat{R},A_1f(\bs{X})\}.$$ 

Next, we show that $||\widehat{f}^*_n||$ is bounded. For any $f\in\overbar{\mc{F}},$ 
$$\mathbb{P}_n\widehat{L}_{\psi}(\widehat{f}^*_n)+\lambda_n||\widehat{f}_n^*||^2\leq\mathbb{P}_n\widehat{L}_{\psi}(f)+\lambda_n||f||^2,$$ by definition of $\widehat{f}^*_n.$ If we choose $f=0$, then, for all $n$ large enough,
\begin{align*}
\mathbb{P}_n\widehat{L}_{\psi}(\widehat{f}^*_n)+\lambda_n||\widehat{f}_n^*||^2 &\leq\mathbb{P}_n\widehat{L}_{\psi}(0)+\lambda_n||0||^2\\
&=\mathbb{P}_n\left\{\frac{|\widehat{R}_i|}{P(A_{i,1}|\bs{X}_i)}\right\}\\
&\leq\pi_0^{-1}(2M_0+M_1),
\end{align*}
where the last inequality holds because of Assumptions (1), (5), and (6). Define $M=\pi_0^{-1}(2M_0+M_1)<\infty.$ Then, because $\mathbb{P}_n\widehat{L}_{\psi}(\widehat{f}^*_n)\geq0,$ we have that $\lambda_n||\widehat{f}^*_n||^2\leq M.$

For any bounded $f$, such as $\widehat{f}^*_n$, we may show that $\left|\mathbb{P}_n\left\{L_{\psi}(f)-\widehat{L}_{\psi}(f)\right\}\right|=o_P(1):$

\begin{align*}
    \left|\mathbb{P}_n\left\{L_{\psi}(f)-\widehat{L}_{\psi}(f)\right\}\right|&\leq\mathbb{P}\left|L_{\psi}(f)-\widehat{L}_{\psi}(f)\right|+o_P(1)\\
    &\leq\pi_0^{-1}\mathbb{P}\left||R|\psi\{R,A_1f(\bs{X})\}-|\widehat{R}|\psi\left\{\widehat{R},A_1f(\bs{X})\right\}\right|+o_P(1)\\
    &\leq\pi_0^{-1}\mathbb{P}\left|\max\{|R|,|\widehat{R}|\}\left[\psi\{R,A_1f(\bs{X})\}-\psi\left\{\widehat{R},A_1f(\bs{X})\right\}\right]\right|+o_P(1)\\
    &\leq\pi_0^{-1}(2M_0+M_1)\mathbb{P}\left|\psi\{R,A_1f(\bs{X})\}-\psi\left\{\widehat{R},A_1f(\bs{X})\right\}\right|+o_P(1)\\
    &\leq2M(\pi_0\lambda_n)^{-1}(2M_0+M_1)\mathbb{P}\left|1\{\sign[R]\neq\sign[\widehat{R}]\}\right|+o_P(1)\\
    &=2M^2\lambda_n^{-1}o_P(\lambda_n)+o_P(1)\\
    &=o_P(1)
\end{align*}

Next, we have
\begin{align*}
    \mathbb{P}_nL_{\psi}(\widehat{f}^*_n)&=\mathbb{P}_nL_{\psi}(\widehat{f}^*_n)+\mathbb{P}_n\widehat{L}_{\psi}(\widehat{f}^*_n)-\mathbb{P}_n\widehat{L}_{\psi}(\widehat{f}^*_n)\\
    &\leq\mathbb{P}_{n}\widehat{L}_{\psi}(\widehat{f}^*_n)+\left|\mathbb{P}_n\left\{L_{\psi}(\widehat{f}^*_n)-\widehat{L}_{\psi}(\widehat{f}^*_n)\right\}\right| + \lambda_n||\widehat{f}^*_n||^2\\
    &\leq\mathbb{P}_n\widehat{L}_{\psi}(f^*)+\lambda_n||f^*||^2+\left|\mathbb{P}_n\left\{L_{\psi}(\widehat{f}^*_n)-\widehat{L}_{\psi}(\widehat{f}^*_n)\right\}\right|\\
    &\leq\mathbb{P}_n{L}_{\psi}(f^*)+\lambda_n||f^*||^2+\left|\mathbb{P}_n\left\{L_{\psi}(\widehat{f}^*_n)-\widehat{L}_{\psi}(\widehat{f}^*_n)\right\}\right|+\left|\mathbb{P}_n\left\{L_{\psi}(f^*)-\widehat{L}_{\psi}(f^*)\right\}\right|.
\end{align*}
Taking the $\limsup$ on both sides, we find $$\limsup_{n\to\infty}\mathbb{P}_nL_{\psi}(\widehat{f}^*_n)\leq\mathbb{P}L_{\psi}(f^*)+o_P(\lambda_n)\leq\mathbb{P}L_{\psi}(\widehat{f}^*_n)+o_P(\lambda_n)$$

Thus, it suffices to show that $\mathbb{P}_nL_{\psi}(\widehat{f}^*_n)-\mathbb{P}_n(L_{\psi}(\widehat{f}^*_n)\to_P0$. Because $\lambda_n||\widehat{f}^*_n||^2$ is bounded by $M$, $\{\sqrt{\lambda_n}f:||\sqrt{\lambda_n}f||\leq\sqrt{M}\}$ is contained in a Donsker class. Note that $\psi(u,v)$ is Lipschitz continuous with respect to $v$, and, thus, $L_{\psi}(f)$ is Lipschitz continuous with respect to $f$. Therefore, $\{\sqrt{\lambda_n}L_{\psi}(f):||\sqrt{\lambda_n}f||\leq\sqrt{M}\}$ is also Donsker. This gives us $\sqrt{n\lambda_n}\{\mathbb{P}_n-\mathbb{P}\}L_{\psi}(\widehat{f}^*_n)=O_p(1)$, which implies $\{\mathbb{P}_n-\mathbb{P}\}L_{\psi}(\widehat{f}^*_n)=o_P(1).$ We finally arrive at $\left|\mc{R}_{\psi}(f^*)-\mc{R}_{\psi}(\widehat{f}^*_n)\right|=o_P(1)$. Furthermore, when $f^*_0\in\overbar{\mc{F}},$ $f^*_0=f^*$, and $$\left|\mc{R}(\widehat{f}^*_n)-\mc{R}(f_0)\right|\leq\left|\mc{R}_{\psi}(\widehat{f}^*_n)-\mc{R}_{\psi}(f^*_0)\right|=o_P(1),$$ where the first inequality holds from \cite{bartlett2006convexity}.

\end{document}